\documentclass[12pt,twoside]{article}
\usepackage[mathscr]{eucal}
\usepackage{amsmath,amsfonts,amssymb,amsthm,mathabx,empheq}
\usepackage{times}
\usepackage{pdfsync}
\usepackage{cite}
\usepackage{url}
\usepackage{hyperref}
\usepackage{tensor}
\usepackage{color}
\usepackage{multicol}
\usepackage{bbold}

\advance\textheight\topskip
\allowdisplaybreaks[1]

\newcommand\td{\text{d}}
\newcommand\cO{{\cal O}}

\newcommand{\p}{\partial}

\newcommand{\be}{\begin{equation}}
\newcommand{\ee}{\end{equation}}
\newcommand{\bea}{\begin{eqnarray}}
\newcommand{\eea}{\end{eqnarray}}

\def\bz{\bar z}

\def\bp{\bar \partial}
\def\n{\nabla}

\def\bn {\overline\nabla}

\newcommand*\xbar[1]{%
  \hbox{%
    \vbox{%
      \hrule height 0.5pt 
      \kern0.3ex
      \hbox{%
        \kern-0.0em
        \ensuremath{#1}%
        \kern-0.0em
      }%
    }%
  }%
}

\allowdisplaybreaks[1]

\DeclareFontFamily{OT1}{rsfs}{} \DeclareFontShape{OT1}{rsfs}{m}{n}{
<-7> rsfs5 <7-10> rsfs7 <10-> rsfs10}{}
\DeclareMathAlphabet{\mycal}{OT1}{rsfs}{m}{n}

\hypersetup{
colorlinks=true,
linktoc=page,    
linkcolor=blue,
citecolor=blue,
urlcolor=blue,
colorlinks=true,
}

\begin{document}

\title{Geometric interpretation of electromagnetic memory and large gauge transformation}

\author{Pujian Mao}
\date{}

\def\mytitle{Geometric interpretation of electromagnetic memory and large gauge transformation}

\addtolength{\headsep}{4pt}

\begin{centering}

  \vspace{1cm}

  \textbf{\Large{\mytitle}}

  \vspace{1cm}

  {\large Pujian Mao }

\vspace{.5cm}

\vspace{.5cm}
\begin{minipage}{.9\textwidth}\small \it  \begin{center}
     Center for Joint Quantum Studies, Department of Physics,\\
     School of Science, Tianjin University, 135 Yaguan Road, Tianjin 300350, China
 \end{center}
\end{minipage}

\end{centering}

\begin{center}
Email: pjmao@tju.edu.cn
\end{center}

\begin{center}
\begin{minipage}{.9\textwidth}
\textsc{Abstract}: We develop a geometric characterization of electromagnetic memory in terms of the optical data of charged-particle congruences. We derive the worldline deviation of charged particles near null infinity for both massless and massive particles and construct the corresponding optical data. We show that the asymptotic shear of the charged-particle congruence provides an equivalent encoding of electromagnetic memory and its associated large gauge transformation. This establishes a geometric interpretation of electromagnetic memory and large gauge transformations in terms of the asymptotic evolution of charged-particle congruences.

\end{minipage}
\end{center}

\thispagestyle{empty}


\section{Introduction}

Gravitational memory is the non-oscillatory contribution to gravitational waves \cite{Zeldovich:1974gvh,Braginsky:1986ia,1987Natur}, which is a generic feature of gravitational radiation at the nonlinear level \cite{Christodoulou:1991cr,Wiseman:1991ss,Thorne:1992sdb}. The passage of gravitational radiation through a region induces a transition from one gravitational vacuum to another. The different vacua are related by asymptotic symmetry transformations of the gravitational theory \cite{Strominger:2014pwa,Pasterski:2015tva}, establishing a concrete equivalence between gravitational memory and asymptotic symmetry \cite{Strominger:2017zoo}.

An analogous structure arises in gauge theory. In particular, electromagnetic (EM) memory \cite{Bieri:2013hqa} provides the gauge-theoretic counterpart of gravitational memory and is related to large gauge transformations (LGTs) \cite{Pasterski:2015zua}. The electromagnetic case offers a considerably simpler setting to study the interplay between memory effects and asymptotic symmetries.

Gravitational memory can be understood as the cumulative effect of geodesic deviation induced by gravitational radiation. In particular, it is determined by the permanent change of the asymptotic shear of null congruence approaching null infinity \cite{Christodoulou:1991cr,Frauendiener:1992dmu,Bieri:2013ada,Tolish:2014bka} and certain time integrals of the asymptotic shear \cite{Pasterski:2015tva,Nichols:2017rqr,Nichols:2018qac,Mao:2018xcw,Compere:2019odm,Mao:2020vgh,Seraj:2021rxd,Seraj:2022qyt,Seraj:2022qqj}. While the observables appear fundamentally different for EM memory. It is the cumulative effect within the radiative transverse vector potential \cite{Bieri:2013hqa} and its time integration \cite{Mao:2017axa,Mao:2019sph,Mao:2021eor,Seraj:2022qqj}. The natural memory effects of EM theory are the velocity kick \cite{Bieri:2013hqa} and the displacement \cite{Mao:2017axa,Mao:2019sph,Mao:2021eor,Seraj:2022qqj} of single moving charged particle.

The change in the motion of a single charged particle provides a direct observational signature of the EM memory, but it is a local response to the EM field and, by itself, does not directly encode the change in asymptotic field data that defines the EM memory. In the gravitational case, memory is instead characterized intrinsically by a permanent change in asymptotic geometric data of the spacetime. This distinction may also help explain why EM memory has received much less attention from the perspective of direct detection \cite{Susskind:2015hpa,Bieri:2023btq,Zosso:2025orw,Zosso:2026aar}, despite the fact that it should be much easier to measure than gravitational memory. This motivates seeking an analogous characterization of EM memory in terms of the optical data of a congruence of charged test particles \cite{FilipeCosta:2006fz}.

In this paper, we derive the worldline deviation of charged particles in both the massless and massive cases. The tangent vectors of the corresponding worldlines are obtained as asymptotic expansions near null infinity. The optical data of the associated charged-particle congruences are constructed. We show that the asymptotic shear of the congruence, for both massless and massive particles, provides an equivalent encoding of the electromagnetic memory information. In particular, it furnishes a previously unexplored geometric observable whose permanent change encodes the same memory information and transforms under LGTs in a manner analogous to the transformation of asymptotic shear under BMS transformations in gravity. This places EM memory and LGTs within the same geometric framework that underlies gravitational memory and asymptotic symmetries.

Throughout the analysis, we neglect the mutual interactions among the charged particles and work in the test-particle approximation. Although such interactions can be incorporated independently and modify the detailed evolution of the congruence, the identification of memory with asymptotic optical data does not rely on the detailed dynamics of these interactions.

To motivate the electromagnetic construction, we first consider linearized gravity as a prototype example. The gravitational analysis makes transparent the relation between memory and the optical data of a null geodesic congruence near null infinity and provides a natural geometric framework for the subsequent electromagnetic discussion.


\section{Null geodesics near null infinity}

We study linearized theory on Minkowski background where the full spacetime line-element is given by perturbative expansions
\be\label{grplane}
\td s^2= \td s^2_{\text{M}} + G h_{\mu\nu} \td x^\mu \td x^\nu + \cO(G^2). 
\ee
We work in flat null coordinates $(u,r,z,\bz)$ \cite{He:2019jjk} and the Minkowski background line-element is given by
\be\label{plane}
\td s^2_{\text{M}}=- 2 \td u \td r  +  2r^2 \td z \td \bz,
\ee
which yields much simpler nonzero connection coefficients
\be
\Gamma^u_{z\bz}=r,\qquad \Gamma^z_{rz}=\frac1r.
\ee
In this case, the topology of the null infinity is $\mathbb{R}\times \mathbb{C}$.

We apply the Newman-Unti (NU) gauge \cite{Newman:1962cia} for the perturbative metric at the linear order $h_{r\mu}=0$ \cite{Conde:2016rom}. The source-free Pauli-Fierz equations yield the following solutions in series expansion near null infinity:
\be
\begin{split}
&h_{zz}=C_{zz} r + D_{zz} + \frac{h_{1zz}}{r}  + \cO(\frac{1}{r^2}),\qquad h_{z\bz}=0,\\ 
&h_{uu}=\frac{M}{r} - \frac{1}{2r^2} (\bp N_z + \p N_z) + \frac{1}{4r^3} (\bp^2 h_{1zz} + \p^2 h_{1\bz\bz}) + \cO(\frac{1}{r^4}),\\
&h_{uz}=\frac12 \bp C_zz + \frac{N_z}{r} - \frac{3}{4r^2} \bp h_{1zz} + \cO(\frac{1}{r^3}),\\
& \p_u M=\frac12 \p_u (\bp^2 C_{zz} + \p^2 C_{\bz\bz}),\qquad \p_u N_z = \frac13 \p M + \frac16 \p^3 C_{\bz\bz} - \frac16 \bp^2 \p C_{zz},\\
& \p_u D_{zz}=0=\bp D_{zz},\qquad \p_u h_{1zz}=\frac12 \p N_z,
\end{split}
\ee
where $\p=\frac{\p}{\p z}$ and $\bp=\frac{\p}{\p \bz}$. In the solution space, $C_{AB}$ is the asymptotic shear tensor for the null geodesic congruence generated by null vector $\frac{\p}{\p r}$, where $A=(z,\bz)$. There is no constraint from Pauli-Fierz equations on the time evolution of the shear tensor. Gravitational memory is fully characterized by the shear tensor $C_{AB}$ and its certain retarded time integrals \cite{Christodoulou:1991cr,Frauendiener:1992dmu,Bieri:2013ada,Tolish:2014bka,Pasterski:2015tva,Nichols:2017rqr,Nichols:2018qac,Mao:2018xcw,Compere:2019odm,Mao:2020vgh,Seraj:2021rxd,Seraj:2022qyt,Seraj:2022qqj}. Indeed, the memory effect in vacuum linearized gravity arises at order $(G^2)$. The aim of the computation in this section is to establish a direct connection between memory and optical data. Such a connection is particularly valuable for the electromagnetic case, where memory is conventionally characterized by a velocity kick, whereas in gravity the effect admits a natural geometric interpretation.

The NU gauge is based on null foliation of spacetime, where $\frac{\p}{\p r}$ is the tangent vector of the null hypersurface and characterizes the propagation of gravitational radiation. The corresponding asymptotic shear tensor natural captures the information of gravitational waves. However, there is no such structure in EM theory. Nevertheless, one can investigate null geodesics along other directions near null infinity to recover gravitational memory, see, e.g., \cite{Mao:2018xcw} for relevant constructions.

Consider a null vector $V$ which is tangent to null geodesic. The geodesic equation up to order $G$ reads
\be
V^\mu \bn_\mu V^\nu + G \xbar\Gamma^\nu_{\alpha\beta} V^\alpha V^\beta =0,
\ee
where $\bn_\mu$ is the covariant derivative associated to flat background metric \eqref{plane}, and $ \xbar\Gamma^\nu_{\alpha\beta}$ is the order $G$ connection. We assume that the null vector can be expanded near null infinity as
\be
V^u=1+\sum_{i=1} \frac{V_i^u}{r^i}, \qquad V^r=\sum_{i=1} \frac{V_i^r}{r^i}, \qquad
V^z= \sum_{i=2} \frac{V_i^z}{r^i}.
\ee
Those asymptotic behaviors guarantee that $V^2=0$ at $r\to\infty$, hence $V^2=0$ on the full trajectory as it is conserved along the geodesic.

Then we solve the geodesic equation in series expansion, and the solutions to relevant leading orders are
\begin{align}
& V_1^u=0,\qquad V_1^r=\frac12 G M,\\
&V_2^z=-\frac12 G \p C_{\bz\bz},\quad V_2^u=\frac12 G \int M \td u' , \quad V_2^r=-\frac14 G (\p N_{\bz} + \bp N_z) ,\\
&V_3^z=- G N_{\bz} + \frac12 G \int \bp M \td u', \qquad V_3^u= -\frac12 G \int (\p N_{\bz} + \bp N_{z} ) \td u'.
\end{align}
We set all initial data of the null vector to zero, as nonvanishing initial data lead to divergent contributions as $u\to \infty$. Accordingly, the massless particle initially moves along a fix-$r$ hypersurface. The gravitational radiation induces subleading corrections to the trajectory. We drop order $G^2$ terms, since the metric is constructed only up to order $G$.

Now we compute the geodesic derivation. For null congruence, we need to introduce another null vector, $k$, as the displacement direction, which satisfied $k\cdot V=-1$. The induced co-dimension two metric is defined as $q_{\mu\nu}=g_{\mu\nu} + V_\mu k_\nu + k_\mu V_\nu$. Then the deviation tensor of $V$ is obtained from the induced metric as $B_{\mu\nu}^V=q^\alpha_\mu q^\beta_\nu \n_\alpha V_\beta$. We choose
\be
\begin{split}
&k_u=-1+\frac{V_1^u}{r}+\frac{V_2^u-(V_1^u)^2}{r^2}+\frac{V_3^u - 2 V_1^u V_2^u + (V_1^u)^3}{r^3} + \cO(\frac{1}{r^4}),\\
&k_r=0,\qquad k_z=0.
\end{split}
\ee
Finally, we obtain the optical data for the null congruence as: the expansion 
\be
\rho=-\frac{G}{2r^2} (\bp^2 C_{zz} + \p^2 C_{\bz\bz} - 2M ) + \cO(\frac{1}{r^3}),
\ee
the transversal shear
\be\label{sigma}
\sigma_{zz}=\frac12 G \p_u C_{zz} r - \frac12 G  C_{zz} + \cO(\frac{1}{r}),
\ee
and twist
\be
\omega_{z\bz}=\cO(\frac{1}{r^2}).
\ee

Gravitational memory is encoded in the permanent change of the asymptotic shear, $\Delta C_{AB}=C_{AB}(u_f) - C_{AB}(u_i)$. The equivalence between gravitational memory and asymptotic symmetry is realized by the fact that the initial and final asymptotic configurations of the shear tensor are related by a BMS transformation \cite{Strominger:2014pwa,Pasterski:2015tva}. Interestingly, the asymptotic shear of the null congruence associated with the null vector $V^\mu$ contains the same memory information. Although the leading term in \eqref{sigma} is given by the news tensor, $N_{AB}=\p_u C_{AB}$, the news vanishes for stationary configuration. Consequently, the permanent change in the asymptotic shear is determined by the difference of the asymptotic values of $C_{AB}$, which therefore provides the leading contribution to the corresponding change in the asymptotic shear. Note that we have omitted terms of order $G^2$, since the metric is constructed only up to order $G$. In the next section, we show that an analogous worldline formulation for charged massless and massive particles similarly encodes the EM memory.


\section{Charged particle worldline near null infinity and LGT}

In this section, we study the worldline of charged particles under electromagnetic fields in flat spacetime. We work again in flat null coordinates \eqref{plane}. We adopt the retarded radial gauge for the gauge fields \cite{He:2014cra}
\be
A_r=0,\qquad A_u=\cO(r^{-1}).
\ee
In this gauge, large gauge transformations are generated by an arbitrary function $\epsilon(z,\bz)$ on the celestial plane, which yields residual gauge transformations $\delta_\epsilon A_\mu=\p_\mu \epsilon(z,\bz)$.

The asymptotic expansion of solutions to Maxwell's equations coupled to a generic source takes the form 
\be
\begin{split}
&J_u=\frac{J_{0u}}{r^2} + \frac{J_{1u}}{r^3} + \cO(\frac{1}{r^4}),\qquad J_r=\frac{J_{0r}}{r^4} + \frac{J_{1r}}{r^5} + \cO(\frac{1}{r^6}),\\
& J_z=\frac{J_{0z}}{r^2} + \frac{J_{1z}}{r^3} + \cO(\frac{1}{r^4}),\qquad A_z=A_{0z} + \frac{A_{1z}}{r} + \frac{A_{2z}}{r^2} + \cO(\frac{1}{r^3}),\\
&A_u=\frac{Q}{r} - \frac{1}{2r^2} (\p A_{1\bz} + \bp A_{1z} - 2 J_{0r}) + \cO(\frac{1}{r^3}),\\
&\p_u Q=\p_u (\bp A_{0z} + \p A_{0\bz}) + J_{0u},\qquad \p_u A_{1z}=\frac12 (\p^2 A_{0\bz} - \p \bp A_{0z} + \p Q + J_{0z}).
\end{split}
\ee
To specify EM memory from these asymptotic data, we consider the motion of charged test particles in the corresponding EM fields. The tangent vector $V^\mu$ of the worldline of a particle with charge $q$ satisfies
\be
V^\mu \bn_\mu V^\nu + q V^\mu {F_\mu}^\nu=0, \qquad F_{\mu\nu}=\p_\mu A_\nu - \p_\nu A_\mu.
\ee
In the following subsections, we investigate both null $(V^2=0)$ and timelike $(V^2=-1)$ worldlines and show how their asymptotic behavior encodes the EM memory.

\subsection{Massless case}

For massless charged particles, we assume that the tangent vector $V^\mu$ admits the asymptotic expansion near null infinity
\be
V^u=1+\sum_{i=1} \frac{V_i^u}{r^i}, \qquad V^r=\sum_{i=1} \frac{V_i^r}{r^i}, \qquad
V^z= \sum_{i=2} \frac{V_i^z}{r^i},
\ee
which indicates that $V=\td r$ at null infinity, namely it is moving along null infinity. Electromagnetic radiation propagates along the other null direction and finally reaches null infinity, where it interacts with the charged particle. Since $V^2$ is conserved along the worldline and $V^2=0$ at $r\to\infty$, the trajectory remains null everywhere.

Solving the worldline equation order by order yields
\be\label{kick}
\begin{split}
&V_1^u=0=V_1^r,\qquad V_2^z=-q A_{0\bz},\qquad V_2^u=q \int Q \td u' , \qquad V_2^r=q^2 A_{0z}A_{0\bz},\\
&V_3^z=-q A_{1\bz} + q \int \bp Q \td u', \qquad V_3^u= -q \int (\p A_{1\bz} + \bp A_{1z} + 2 q A_{0z}A_{0\bz}) \td u'.
\end{split}
\ee
We set all integration constants to zero, as they would lead to divergences in the $u\to\infty$ limit. In the absence of electromagnetic radiation, the worldline follows the null generator of $V=\td r$ on constant-$r$ hypersurface. The electromagnetic field induces subleading corrections to the trajectory, hence generate both transverse and radial motion.

We now compute the worldline deviation and the associated optical data. Introduce an auxiliary null vector $k^\mu$, normalized such that $k\cdot V=-1$, with asymptotic forms
\be
\begin{split}
&k_u=-1+\frac{V_1^u}{r}+\frac{V_2^u-(V_1^u)^2}{r^2}+\frac{V_3^u - 2 V_1^u V_2^u + (V_1^u)^3}{r^3} + \cO(\frac{1}{r^4}),\\
&k_r=0,\qquad k_z=0.
\end{split}
\ee
The optical expansion of the congruence of charged particles is
\be
\rho=-\frac{q}{r^2} (\bp A_{0z} + \p A_{0\bz}) + \cO(\frac{1}{r^3}),
\ee
the shear is
\be
\sigma_{zz}=-q \p A_{0z} + \cO(\frac{1}{r}), \qquad \sigma_{z\bz}= \cO(\frac{1}{r}),
\ee
and the twist is
\be
\omega_{z\bz}=\frac12 q (\p A_{0\bz} - \bp A_{0z}) + \cO(\frac{1}{r}).
\ee

The EM memory is encoded in the permanent change of the radiative data, $\Delta A_{0z(\bz)}=A_{0z(\bz)}(u_f) - A_{0z(\bz)}(u_i)$, which induces the velocity kick of a charged particle in \eqref{kick} \cite{Bieri:2013hqa}. The equivalence between EM memory and LGT is realized by the fact that the initial and final gauge potentials are related by a LGT \cite{Pasterski:2015zua}. The explicit expressions for the optical data derived above show that they are determined by the same radiative data $A_{0z(\bz)}$. Consequently, the permanent change in the optical data of the charged-particle congruence encodes precisely the same memory information as $\Delta A_{0z(\bz)}$, which provides a geometric characterization of the EM memory effect and the corresponding LGT. Consider a LGT generated by $\epsilon(z,\bar z)$. Since $\delta_\epsilon A_{0z(\bz)}=\p_{z(\bz)} \epsilon$, the optical data transform as
\be
\begin{split}
&\delta_\epsilon\rho=-\frac{2q \bp \p \epsilon}{r^2}  + \cO(\frac{1}{r^3}),\qquad \delta_\epsilon \sigma_{zz}=-q \p^2 \epsilon + \cO(\frac{1}{r}), \\
&\delta_\epsilon \sigma_{z\bz}= \cO(\frac{1}{r}),\qquad \delta_\epsilon \omega_{z\bz}= \cO(\frac{1}{r}).
\end{split}
\ee

In these computations, we have neglected the interactions among the charged particles in the congruence. Such interactions can be incorporated independently and modify the detailed evolution of the congruence. However, the memory effect considered here is encoded in the asymptotic radiative data $A_{0z(\bz)}$ and the associated LGT relating the initial and final configurations. Consequently, the identification of EM memory with the permanent change of the optical data is insensitive to the mutual interactions of the particles and is determined entirely by the radiative sector.

\subsection{Massive case}

In this subsection, we present the parallel computation for massive charged particles. We assume that the tangent vector $V^\mu$ admits the asymptotic expansion 
\be
V^u=1+\sum_{i=1} \frac{V_i^u}{r^i}, \qquad V^r=\frac12 + \sum_{i=1} \frac{V_i^r}{r^i}, \qquad
V^z= \sum_{i=2} \frac{V_i^z}{r^i},
\ee
which indicates that the massive charged particles asymptotically follow the timelike curves on the hypersurfaces $r-\frac12 u=\mathrm{const.}$ at large radial distance. Electromagnetic radiation propagates across these hypersurfaces and interacts with the charged particle. Since $V^2$ is conserved along the worldline and $V^2=-1$ at $r\to\infty$, the trajectory remains timelike everywhere.

Solving the worldline equation for the massive charged particle order by order yields
\be
\begin{split}
&V_1^u=0=V_1^r,\qquad V_2^z=-q A_{0\bz}, \qquad V_2^u=q \int Q \td u' , \\
& V_2^r=q^2 A_{0z}A_{0\bz} - \frac12 q \int Q \td u',\qquad
V_3^z=-q A_{1\bz} + q \int \bp Q \td u' .
\end{split}
\ee
We again set all integration constants to zero. In the absence of electromagnetic radiation, the worldline follows a timelike curve on constant-$(r-\frac12 u)$ hypersurface. The electromagnetic field induces subleading corrections to the trajectory. 

We now compute the worldline deviation for the congruence of massive charged particles. We define the projection tensor
\be
P^\mu_\nu=\delta^\mu_\nu+V^\mu V_\nu,\qquad P^\mu_\nu V^\nu=0=P^\mu_\nu V_\mu,
\ee
from which we define the deviation tensor 
\be
B_{\mu\nu}=\n_\alpha V_\beta P^\alpha_\mu P^\beta_\nu. 
\ee
The corresponding expansion is 
\be
\rho=\frac2r-\frac{2q}{r^2} (\bp A_{0z} + \p A_{0\bz} - Q) + \cO(\frac{1}{r^3}),
\ee
the relevant component of the shear is
\be
\sigma_{zz}=- 2 q \p A_{0z} + 2q^2 A_{0z} \p_u A_{0z}  + \cO(\frac{1}{r}),
\ee
and relevant component of the twist is
\be
\omega_{z\bz}=\frac12 q (\p A_{0\bz} - \bp A_{0z}) + \frac{q^2}{2} (A_{0\bz}\p_u A_{0z} - A_{0z} \p_u A_{0\bz}) + \cO(\frac{1}{r}).
\ee

The permanent change of the radiative data induces velocity kick of a massive charged particle \cite{Bieri:2013hqa}. The optical data derived above show that, for stationary configurations $\p_u A_{0z(\bz)}=0$, the asymptotic shear and twist are completely determined by the same radiative data $A_{0z(\bz)}$ that encode the electromagnetic memory effect, whereas the expansion also depends on the Coulombic sector through the charge aspect $Q$. Nevertheless, the transformation law of the optical data for the massive case is the same as the massless case under LGT. Consequently, the optical data of the massive charged-particle congruence provide a geometric characterization of EM memory analogous to that obtained in the massless case.


\section{Discussions} 

In this work, we developed a geometric characterization of electromagnetic memory in terms of the optical data of charged-particle congruences. We derived the worldline deviation and asymptotic optical data for both massless and massive charged particles, and showed that their asymptotic shear provides an equivalent encoding of the electromagnetic memory. Since the initial and final radiative data are related by a large gauge transformation, this result gives a geometric interpretation of the relation between EM memory and LGTs in terms of the asymptotic evolution of charged-particle congruences.

An interesting direction for future work is to apply this framework to binary neutron star mergers \cite{Bamber:2025ggq,Brabant:2026usw}, where gravitational and electromagnetic memory can be studied within a common geometric framework and compared on the same footing. Another interesting direction is to explore whether the correspondence between gravitational and electromagnetic memory through congruence optical data can be understood within the framework of the classical double copy \cite{Gonzo:2021drq}.


\section*{Acknowledgments} 

This work is supported in part by the National Natural Science Foundation of China (NSFC) under Grants No.~12475059 and No.~11935009, and by Tianjin University Self-Innovation Fund Extreme Basic Research Project Grant No. 2025XJ21-0007.


\providecommand{\href}[2]{#2}\begingroup\raggedright\endgroup

\end{document}